\documentclass{aa} 
\usepackage{graphicx}
\usepackage{txfonts}
\usepackage{multirow}
\usepackage{caption, threeparttable}
\usepackage{xcolor}
\usepackage{float}
\usepackage{amsmath}

\begin{document}

\newcommand{\teff}{$T_\mathrm{eff}$}
\newcommand{\logg}{$\log g$}
\newcommand{\feh}{[Fe/H]}
\newcommand{\microturb}{$\xi_\mathrm{micro}$}
\newcommand{\sife}{[Si/Fe]}
\newcommand{\mgfe}{[Mg/Fe]}
\newcommand{\afe}{[$\alpha$/Fe]}
\newcommand{\co}{CO($\nu =2-0$)\,}

\newcommand{\water}{H$_2$O}
\newcommand{\invcm}{cm$^{-1}$}
\newcommand{\kms}{km\,s$^{-1}$}
\newcommand{\mic}{$\mu \mathrm m$}
\newcommand{\msun}{$\mathrm{M}_\odot$}

\defcitealias{Nandakumar:2023_I}{Paper\,I}
\defcitealias{Nandakumar:2023b_II}{Paper\,II}
\defcitealias{Nandakumar:24_III}{Paper\,III}
\titlerunning{Ba identification} 
\authorrunning{Nandakumar et al.}

   \title{M giants with IGRINS
}
   \subtitle{IV. Identification and characterization of a near-IR line of the s-element Barium}

   \author{G. Nandakumar
            \inst{1}
            \and
          N. Ryde
          \inst{1}
          \and
        H. Hartman
          \inst{2}
          \and      
        G. Mace
        \inst{3}
          }

   \institute{Division of Astrophysics, Department of Physics, Lund University, Box 118, SE-221 00 Lund, Sweden\\
              \email{govind.nandakumar@fysik.lu.se}
            \and
    Materials Science and Applied Mathematics, Malm\"o University, SE-205 06 Malm\"o, Sweden
             \and
    Department of Astronomy and McDonald Observatory, The University of Texas, Austin, TX 78712, USA
     }

   \date{Received ; accepted }

 
  \abstract
   {Neutron-capture elements represent an important nucleosynthetic channel in the study of the Galactic Chemical Evolution of stellar populations. For stellar populations behind significant extinction, such as those in the Galactic Center and along the Galactic plane, abundance analyses based on near-infrared (near-IR) spectra are necessary.  Previously, spectral lines from the neutron-capture elements such as copper (Cu), cerium (Ce), neodymium (Nd), and ytterbium (Yb) have been identified in the H band, while yttrium (Y) lines have been identified in the K band.}
   {Due to the scarcity of spectral lines from neutron-capture elements in the near-IR, the addition of useful spectral lines from other neutron-capture elements is highly desirable.  The aim of this work is to identify and characterise a spectral line suitable for abundance determination from the most commonly used s-process element, namely barium.}  
   {We observed near-infrared spectra of 37 M giants in the solar neighborhood at high signal-to-noise and high spectral resolution using the IGRINS spectrometer on the GEMINI South telescope. The full H- and K-bands were recorded simultaneously at $R=45,000$. Using a manual spectral synthesis method, we determined the fundamental stellar parameters for these stars and derived the barium abundance from the Ba line (6s5d $^3$D$_2 \rightarrow$ 6s6p $^3$P$^o_2$) at $\lambda_\mathrm{air}=23\,253.56\,$\AA\ in the K band.}
   {We demonstrate that the Ba line in the K band at 2.33\,\mic\  ($\lambda$23253.56) is useful for abundance analysis from spectra of M giants. The line becomes progressively weaker at higher temperatures and is only useful in M giants and the coolest K giants at supersolar metallicities. }
    {We can now add Ba to the trends of the heavy elements Cu, Zn, Y, Ce, Nd, and Yb, which can be retrieved from high-resolution H- and K-band spectra. This opens up the study of  nucleosynthetic channels, including the s-process and the r-process, in dust-obscured populations. Thus, these elements can be studied for heavily dust-obscured regions of the Galaxy, such as the Galactic Center.} 


   \maketitle
%

\section{Introduction}
\label{sec:intro}
\vspace{-5pt}

This paper is part of a series utilizing observations from the Immersion Grating Infrared Spectrometer (IGRINS) to study M giants. M giants are luminous, cool giant stars that have been underutilized in spectroscopic studies of stellar populations, primarily because K giants are more straightforward to analyze. However, M giants are the only practical targets for investigating stellar populations in regions of the Milky Way obscured by dust, such as the inner Galaxy. In such studies near-infrared (near-IR)  observations are essential due to the significant reddening and extinction.  
Earlier abundance studies have shown the strength of high-resolution, near-IR spectra recorded with IGRINS in general \citep[see][]{afsar:2018,afsar:2019,afsar:2020,montelius:22,afsar:2023,nandakumar:22_P,Nandakumar:24_bp1}.

The ubiquitous presence of titanium oxide (TiO) spectral lines in optical spectra of M giants prevents traditional spectral analyses conducted at optical wavelengths. However, the near-IR region of M giant spectra proves to be valuable for determining stellar parameters and abundances through spectroscopy \citep[see, e.g.,][]{Nandakumar:2023_I}.

In \citet[][paper I]{Nandakumar:2023_I} an iterative method for determining the stellar parameters of M giants with $3400<$\teff$<4000\,$K from high-resolution, H-band spectra and isochrones was presented. This method utilized  temperature-sensitive OH lines in combination with CO and Fe lines. A total of 50 M-giants, observed with IGRINS, were analysed and bench marked. \citet[][paper II]{Nandakumar:2023b_II} demonstrated the potential and accuracy of this method in the detailed investigation of ten HF lines for determining for the Galactic Chemical Evolution of fluorine. Similarly, in \citet[][paper III]{Nandakumar:24_III}, abundance trends versus metallicity for these 50 solar-neighbourhood stars were derived for 21 elements, including the $\alpha$-elements, as well as odd-Z elements, iron-peak elements, and neutron-capture elements. The paper identified the optimal spectral lines to use and the precision achievable for all 21 elements.  This range of elements is crucial for obtaining a comprehensive chemical view of the Milky Way and its components.

 \begin{figure}
  \includegraphics[width=0.45\textwidth]{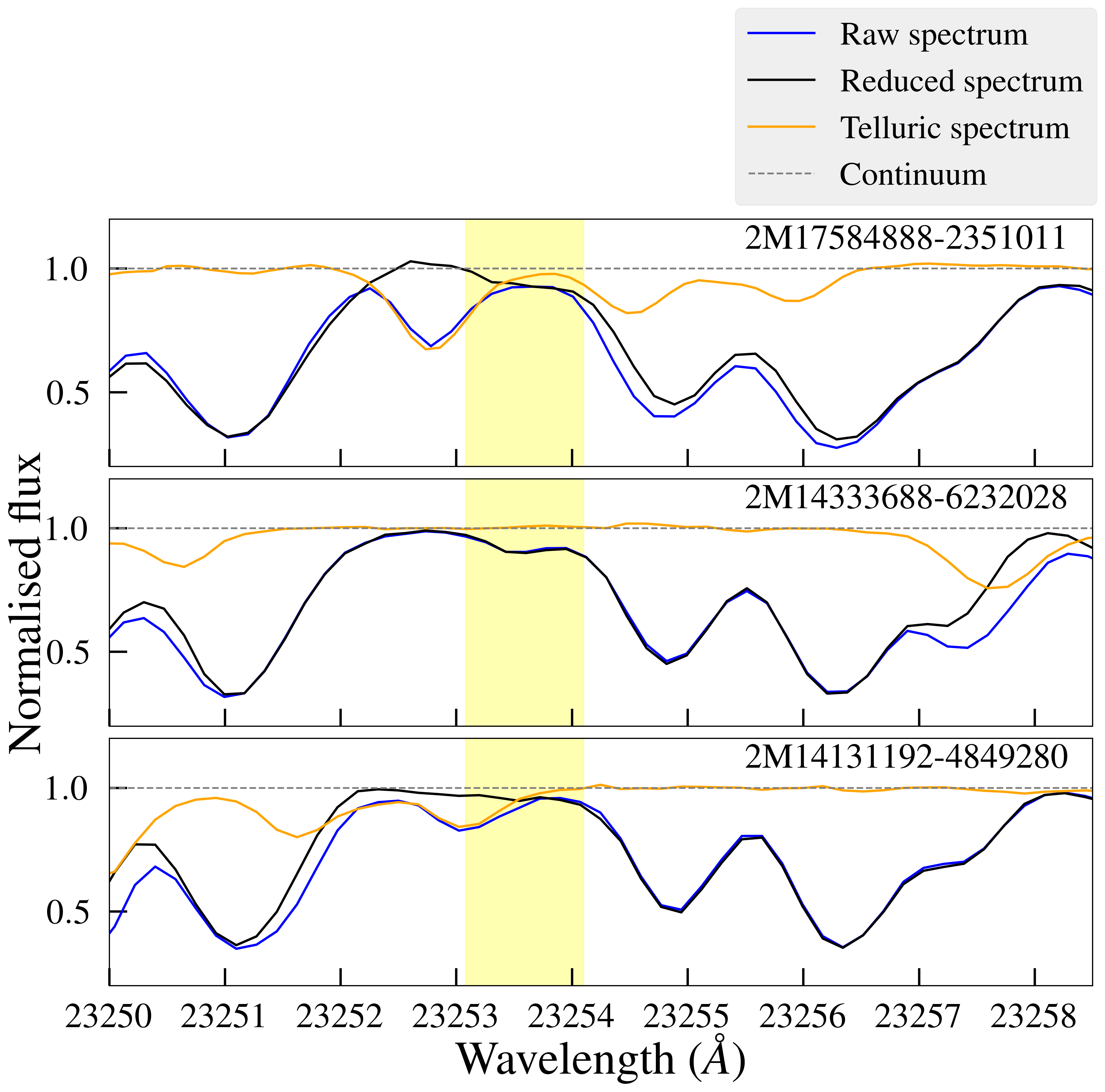}
  \caption{ Example spectra surrounding the barium line (yellow band) of three stars showing the telluric correction carried out in this work. The raw observed target spectra before telluric correction is shown in blue, the telluric corrected reduced spectra in black, and the telluric star spectra in orange. The dashed gray line represent the normalised continuum level at 1.}   
  \label{fig:ba_raw}%
\end{figure}

\begin{table*}
\caption{ Observational details of M giant stars. }\label{table:obs}
\centering
\begin{tabular}{c c c c c c c c}
\hline
\hline
 Index & 2MASS ID  & H$_\mathrm{2MASS}$  & K$_\mathrm{2MASS}$  & Date    & Exposure Time & S/N$_\mathrm{H}$ & S/N$_\mathrm{K}$ \\

 &  &  (mag) & (mag) &  UT &  mm:ss  &  \multicolumn{2}{c}{(per resolution element)} \\
\hline
1 & 2M05484106-0602007  & 7.0 & 6.7 &  2021-01-02  &  06:32   & 110  &  130  \\
2 & 2M05594446-7212111  & 7.0 & 6.8 &  2021-01-02  &  10:35   & 100  &  120   \\
3 & 2M06035214-7255079  & 7.1 & 6.8 &  2021-01-01  &  06:04   & 100  &  110  \\
4 & 2M06052796-0553384  & 7.0 & 6.6 &  2021-01-02  &  07:08   & 100  &  120   \\
5 & 2M06124201-0025095  & 7.0 & 6.7 &  2021-01-01  &  07:34   & 90  &  100   \\
6 & 2M06223443-0443153  & 7.1 & 6.8 &  2021-01-02  &  08:35   & 90  &  110   \\
7 & 2M06231693-0530385  & 7.1 & 6.7 &  2021-01-02  &  08:05   & 80  &  100   \\
8 & 2M06520463-0047080  & 7.1 & 6.7 &  2021-01-02  &  06:41   & 80  &  100   \\
9 & 2M06551808-0148080  & 7.1 & 6.8 &  2021-01-02  &  04:10   & 90  &  100   \\
10 & 2M06574070-1231239  & 7.3 & 6.9 &  2021-01-02  &  09:31   & 90  &  110   \\
11 & 2M10430394-4605354  & 7.2 & 6.9 &  2021-02-11  &  06:22   & 120  &  120  \\
12 & 2M12101600-4936072  & 7.2 & 6.9 &  2021-02-19  &  06:28   & 70  &  90    \\
13 & 2M13403516-5040261  & 7.1  & 6.8  &  2021-03-17  &  11:27   & 170  &  170   \\
14 & 2M14131192-4849280  & 7.2  & 6.9  &  2021-03-17  &  09:02   & 150  &  150   \\
15 & 2M14241044-6218367  & 7.3  & 6.8  &  2021-04-17  &  09:38   & 230  &  260   \\
16 & 2M14260433-6219024  & 7.2 & 6.8  &  2021-02-18  & 05:01   & 50  &  70  \\
17 & 2M14261117-6240220  & 7.0 & 6.5  &  2021-02-14  & 08:09   & 80  &  100  \\
18 & 2M14275833-6147534  & 7.0 & 6.5  &  2021-02-18  & 05:19   & 80  &  100   \\
19 & 2M14283733-6257279  & 6.5 & 6.1  &  2021-02-18  & 06:59   & 70  &  90  \\
20 & 2M14291063-6317181  & 7.4 & 6.9  &  2021-02-18  & 05:16   & 80  &  110   \\
21 & 2M14322072-6215506  & 6.7 & 6.4  &  2021-04-13  &  11:40 & 270  &  290  \\
22 & 2M14332869-6211255  & 6.5 & 6.1  &  2021-02-20  &  06:20  & 120  &  140  \\
23 & 2M14333081-6221450  & 7.2 & 6.7  &  2021-03-18  &  10:10  & 210  &  240  \\
24 & 2M14333688-6232028  & 7.2 & 6.8  &  2021-04-24  &  14:24  & 210  &  250  \\
25 & 2M14345114-6225509  & 7.0 & 6.6  &  2021-04-24  &  13:55  & 270  &  300  \\
26 & 2M14371958-6251344  & 6.7 & 6.3  &  2021-04-24  &  09:18  & 240  &  240   \\
27 & 2M15161949+0244516  & 7.2 & 7.0  &  2021-04-24  &  14:01  & 190  &  190  \\
28 & 2M17584888-2351011  & 7.3 & 6.5  &  2021-04-28  &  08:21  & 200  &  290  \\
29 & 2M18103303-1626220  & 7.3 & 6.5  &  2021-04-28  &  08:47  & 230  &  310  \\
30 & 2M18142346-2136410  & 7.1 & 6.6  &  2021-04-28  &  12:19  & 240  &  280  \\
31 & 2M18191551-1726223  & 7.2 & 6.6  &  2021-04-28  &  14:16  & 320  &  380  \\
32 & HD 132813  &  -0.8  &  -1.0  & 2016-02-29   &  02:03  & 479 & 508 \\ 
33 & HD 175588  &  -0.9  &  -1.2  & 2015-02-11   &  02:03  & 760 &  760\\ 
34 & HD 89758  &  -0.1  &  -0.13  &  2015-04-12 &  02:03  & 483 &  484\\ 
35 & HD 224935  &  -0.7  &  -0.9  &  2015-17-11  &  02:03  & 467  & 505 \\ 
36 & HD 101153  &  0.0  &  -0.3  & 2015-07-12  & 02:03  & 346  & 376 \\ 
37 & HD 96360  &  3.2  &  2.8  & 2016-01-03  & 00:30  & 98 & 94 \\ 
\hline
\hline                                 
\end{tabular}
\end{table*}

Especially, the neutron-capture elements introduce important nucleosynthetic channels with their own evolutionary timescales \citep[see, e.g.,][]{manea:23}. Low-mass  Asymptotic Giant Branch (AGB)  stars are responsible
for most of the main s-process, producing elements such as  barium (Ba), yttrium (Y), cerium (Ce), and neodymium (Nd). Due to these stars' lower masses \citep[1.3-3\,\msun;][]{Grisoni:2020} compared to the progenitors of the supernovae type II (SNII)-dominated elements, the s-process elements are formed on a longer time-scale.  In \citetalias{Nandakumar:24_III} the elemental trends versus metallicity were derived for the neutron-capture elements copper (Cu; weak s-process element), 
Y, Ce, and Nd (main s-process elements), as well as Yb (50/50 r/s-process element). These abundances were determined from H-band lines except for the yttrium abundances, which were determined from Y lines (from Y\,{\sc i}) in the K band at 2.2\,\mic.

In general, there has been a lack of useful spectral lines in the near-IR for determining the abundances of neutron-capture elements. In APOGEE spectra ($R\sim 22,500)$, \citet{Cunha:2017} identified and characterized cerium lines (from Ce\,{\sc ii}) in the H band. Similarly, \citet{Hasselquist:2016}
identified and analysed neodymium lines (from Nd\,{\sc ii}) also in APOGEE spectra. By carefully addressing and treating blends,  \citet{Hayes:2022} were additionally able to  determine the chemical evolution for Cu\,{\sc i} and Nd\,{\sc ii} from APOGEE spectra. Using near-IR spectra observed at higher spectral resolution from the IGRINS spectrometer ($R\sim 45,000)$, \citet{montelius:22} succeeded in deriving the abundance trend for ytterbium (Yb) using a Yb\,{\sc ii} line in H band. Ytterbium is particularly interesting because it has the highest contribution from the r-process among these elements.\footnote{The ratio of the s- and r-process contributions in the solar isotopic mixture are s/r(Y)=70/30 [in $\%$] and s/r(Ce)= 83/17, s/r(Nd)=60/40 , and s/r(Yb)=50/50 \citep{bisterzo:14,prantzos:20,kobayashi:20}.} 


 
In this paper, we identify and characterize a barium (Ba) line from Ba\,{\sc i} in IGRINS spectra in the K-band. This adds the most widely used s-process dominated element to the list of neutron-capture elements that can be used in near-IR spectra. Barium is one of the most s-process dominated element with s/r(Ba)= 90/10 \citep{prantzos:20}.

 \begin{table}
\caption{Stellar parameters, C/O ratio, and  [Ba/Fe] values for each star determined in this work. }\label{table:parameters}
\begin{tabular}{c c c c c c c c}
\hline
\hline
 Index & T$_\mathrm{eff}$ & $\log g$  & [Fe/H]  &  $\xi_\mathrm{micro}$ & C/O  & [Ba/Fe] \\
 \hline
  & K & log(cm/s$^{2}$) & dex & Km/s &   & dex\\
  \hline
1   &  3490  &  0.48  &  -0.28  &  2.03  &  0.41  &  0.04 \\  
2   &  3694  &  0.74  &  -0.45  &  1.88  &  0.38  &  0.17  \\  
3   &  3742 &  1.08  &  0.0  &  1.78  &  0.46  &  0.04 \\  
4   &   3677 &  0.92  &  -0.07  &  1.78  &  0.37  & 0.23  \\  
5    &   3583 &  0.42  &  -0.66  &  2.39  &  0.35  &  0.38  \\  
6    &  3521 &  0.4  &  -0.52  &  2.19  &  0.35  &  0.24 \\  
7    &  3484 &  0.32  &  -0.55  &  2.09  &  0.38  &  0.10  \\  
8    &  3581 &  0.67  &  -0.21  &  2.15  &  0.36  &  0.50  \\  
9    &  3606 &  0.52  &  -0.56  &  1.96  &  0.37  &  0.23  \\  
10    &  3561 &  0.56  &  -0.36  &  2.24  &  0.33  &  0.55  \\  
11    &  3568 &  0.96  &  0.25  &  1.83  &  0.43  &  -0.18  \\  
12    &  3539 &  0.5  &  -0.41  &  2.05  &  0.41  &  0.11  \\  
13    &  3528 &  0.61  &  -0.15  &  1.92  &  0.74  &  -0.02  \\ 
14    &  3504  &  0.61  &  -0.08  &  1.81  &  0.79 &  0.01  \\  
15    &  3543 &  0.8  &  0.11  &  1.95  &  0.52  &  0.10  \\  
16    &  3386 &  0.55  &  0.13  &  1.82  &  0.54  &  -0.05  \\  
17    &  3387 &  0.52  &  0.08  &  1.92  &  0.54  &  -0.11  \\  
18    &  3453 &  0.63  &  0.08  &  1.91  &  0.46  &  0.10  \\  
19    &  3465  &  0.62  &  0.04  &  1.83  &  0.46  &  0.5  \\  
20    &  3430 &  0.54  &  0.0  &  1.95  &  0.51  &  -0.09  \\  
21    &  3639 &  0.89  &  -0.0  &  1.76  &  0.47  &  -0.23   \\  
22    &  3664 &  1.11  &  0.23  &  1.99  &  0.54  &  0.01  \\  
23    &  3430 &  0.55  &  0.02  &  1.92  &  0.45  &  0.10  \\  
24    &  3425 &  0.54  &  0.02  &  1.87  &  0.81  &  0.04  \\  
25    &  3442 &  0.68  &  0.18  &  1.85  &  0.62  &  0.07  \\  
26   &  3650 &  0.98  &  0.1  &  1.8  &  0.46  &  0.07  \\  
27    &  3691 &  0.76  &  -0.4  &  1.98  &  0.36  &  -0.01  \\  
28    &  3564 &  0.95  &  0.25  &  2.2  &  0.59  &  -0.16   \\  
29    &  3347 &  0.46  &  0.09  &  1.98  &  0.81  &  0.11  \\  
30    &  3390 &  0.48  &  0.01  &  1.96  &  0.52  &  -0.07  \\  
31    &  3434 &  0.59  &  0.07  &  1.93  &  0.49  &  0.00  \\  
32    &  3457 &  0.44  &  -0.27  &  1.88  &  0.44  &  0.20  \\  
33    &  3484 &  0.6  &  -0.04  &  2.24  &  0.49  &  0.33  \\  
34    &  3807 &  1.15  &  -0.09  &  1.65  &  0.41  &  0.00  \\  
35    &  3529 &  0.64  &  -0.1  &  2.01  &  0.42  &  0.15 \\  
36    &  3438 &  0.51  &  -0.07  &  2.03  &  0.43  &  0.04  \\  
37    &  3459 &  0.5  &  -0.15  &  1.86  &  0.52  &  0.23  \\  
\hline

\hline
\hline
\end{tabular}
 
\end{table}









\begin{figure*}
  \includegraphics[width=\textwidth]{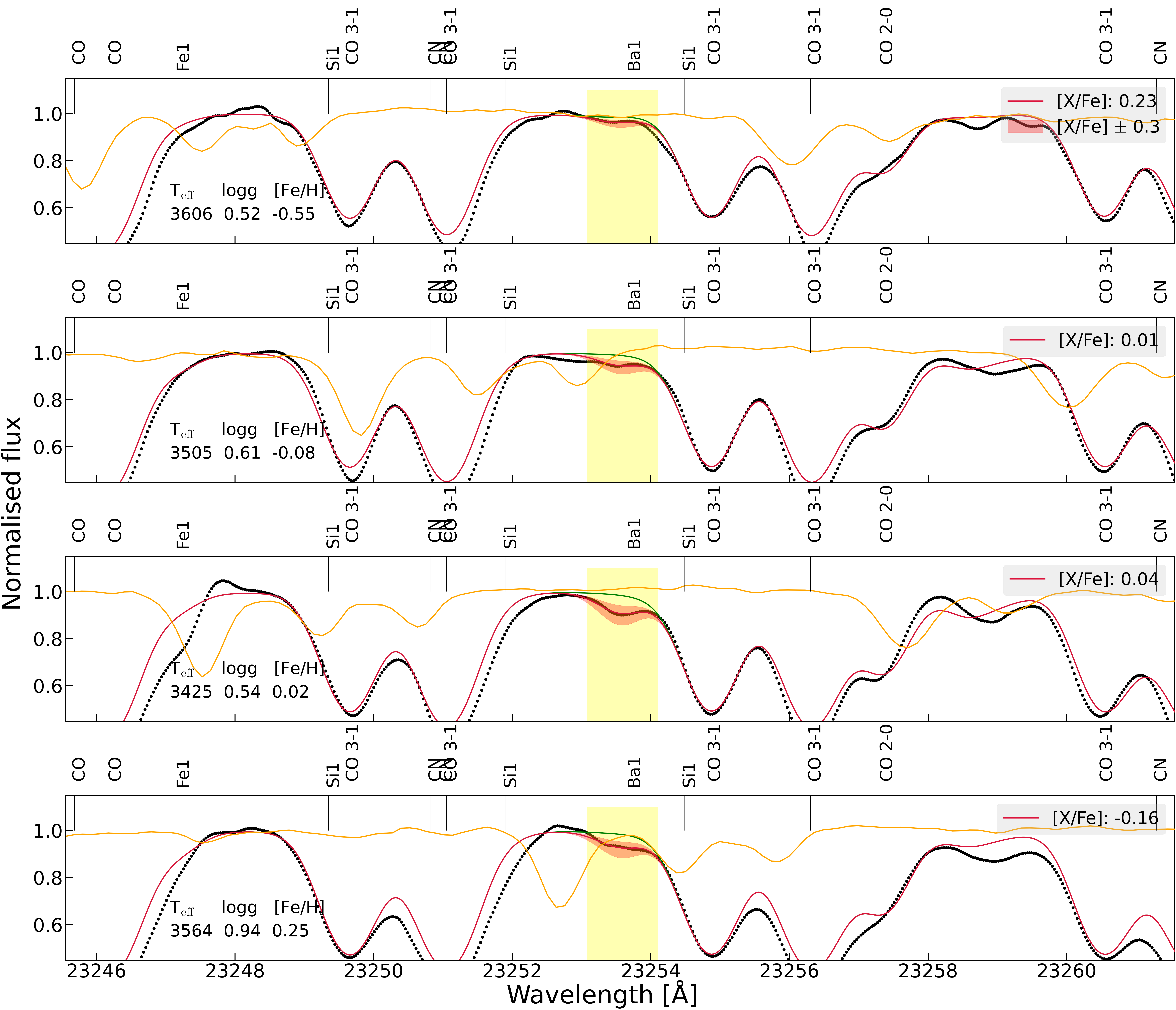}
  \caption{ Wavelength regions centered at the barium line at 23253.56 \AA\, for four stars with \teff\, ranging from 3400 K to 3600 K and \feh\, ranging from -0.5 dex to 0.25 dex. In each panel, the black circles denote the observed spectrum, the crimson line denotes the best-fit synthetic spectrum, and the red band denotes the variation in the synthetic spectrum for a difference of $\pm$0.3\,dex in the [Ba/Fe]. The yellow bands in each panel represent the line masks defined for the Ba line. The green line shows the synthetic spectrum without Ba, and the orange line shows the telluric spectrum of the standard star that is used to correct for telluric contaminations in the observed star spectrum. The [Ba/Fe] value corresponding to the best-fit case for the Ba line are listed in each panel. All identified atomic and molecular lines are also denoted at the top of each panel. }
  \label{fig:ba_spectra}%
\end{figure*}

\section{Observations and data reduction}
\label{sec:obs}

Among the 50 M giants analysed in \citetalias{Nandakumar:2023_I}, we determined the barium (Ba) abundances for 37 
M giants (\teff$< 4000$~K) from high-resolution spectra (spectral resolution of 
$R\sim45,000$) observed with the Immersion GRating INfrared Spectrograph \citep[IGRINS;][]{Yuk:2010, Wang:2010, Gully:2012, Moon:2012, Park:2014, Jeong:2014}. In addition to six nearby M giants available in the IGRINS spectral library \citep{park:18, rrisa}, observed at the McDonald Observatory in 2015 and 2016 \citep{Mcdonald}, the remaining stars were observed with IGRINS mounted on the Gemini South telescope \citep{Mace:2018} under the programs GS-2020B-Q-305 and  GS-2021A-Q302.
These were observed over a period from January to April 2021.
All the stars are located in the solar neighborhood and serve as a good reference sample. 

The IGRINS observations were conducted using one or more ABBA nod sequences along the slit to facilitate sky background subtraction. Exposure times were chosen to achieve an average signal-to-noise ratio (S/N) of at least 100,\footnote{The signal-to-noise ratios were provided by RRISA \citep[The Raw $\&$ Reduced IGRINS Spectral Archive;][]{rrisa} and are the average S/N for the H- and K-bands per resolution element. The S/N varies over the orders and is lowest at the ends of the orders.} resulting in observing times ranging from 1/2 to 15 minutes. While for three of the stars the observations did not reach S/N of 100, for 15 stars the S/N was well above, see Table \ref{table:obs}.

Spectral reductions were performed using the IGRINS Pipeline Package \citep[IGRINS PLP;][]{Lee:2017}. This process included flat-field correction, and A-B frame subtraction, telluric correction, and wavelength calibration using sky OH emission lines \citep{Han:2012, Oh:2014}. To eliminate telluric lines, the science spectra were divided by a telluric standard-star spectrum of a fast-rotating late-B to early-A dwarf, observed close in time and at a similar air mass. Subsequently, the spectral orders of the science targets and telluric standards were stitched together after normalizing each order and then  combining and resampling them in {\tt iraf} \citep{IRAF}, excluding low S/N edges of each order. Finally, the spectra were shifted to laboratory wavelengths in air following stellar radial velocity correction. In Figure \ref{fig:ba_raw} we show the raw observed science spectra before telluric correction (blue line), reduced science spectra after telluric correction (black line), and telluric standard star spectra (orange line) surrounding the barium line of interest in this work (yellow band) for three target stars in this work.   

Considerable effort was invested in defining specific local continua in the spectral segment where the Ba line resides to address any residual modulations in the continuum levels. To make sure we select the best continuum points, we went through the spectrum of each star manually within the 30 \AA\, wavelength window that includes the barium line. We then identified continuum points that are least affected by telluric absorption correction or bad noisy pixels. Thus we used separate continuum mask file for each star while determining barium abundance. Furthermore, thanks to the higher spectral resolution (R $\sim$ 45,000) of IGRINS, we could identify reliable continuum points in between the CO bands in the region. No broad absorption features caused by diffuse interstellar bands \citep[DIBs; see, e.g.,][]{dib:geballe1,dib:geballe_nature} are present near the Ba line or other lines used in this spectral analysis.  Further details on the observations are provided in \citetalias{Nandakumar:2023_I} and \citetalias{Nandakumar:2023b_II}, where stellar parameters and abundance trends versus metallicity for 21 elements for these stars are presented.


\section{Analysis}
\label{sec:analysis} 

The barium feature at 23253.56 \AA\, was analysed in the spectra of the 37 M giants observed, by synthesising spectra for given fundamental stellar parameters. The synthetic spectra were generated using the Spectroscopy Made Easy \citep[SME;][]{sme,sme_code} tool by calculating the spherical radiative transfer through a relevant stellar atmosphere model defined by these fundamental  parameters. We chose the stellar atmosphere model by interpolating in a grid of one-dimensional Model Atmospheres in a Radiative and Convective Scheme (MARCS) stellar atmosphere models \citep{marcs:08}. 



The fundamental stellar parameters, namely, the effective temperature (\teff), surface gravity (\logg), metallicity (\feh), and microturbulence ($\xi_\mathrm{micro}$) were provided in \citetalias{Nandakumar:2023_I} and were determined using the method devised in that paper.  The surface gravity, \logg, is, thus, constrained based on the \teff\ and \feh\ values from 3-10 Gyr Yonsei-Yale isochrones \citep{Demarque:2004}. The stellar parameters and C/O ratios are given in Table \ref{table:parameters} and typical uncertainties are found to be $\pm$ 100 K in \teff\,, $\pm$0.2 dex in \logg\,, $\pm$0.1 dex in \feh\,, and $\pm$0.1 km/s in $\xi_\mathrm{micro}$. For majority of stars, C/O ratios are lower than the solar value ($\sim$ 0.5) indicating they are in the red giant branch (RGB) phase
of the stellar evolution. For the remaining few stars, slightly
higher than solar C/O ratio are found, which might be within uncertainties or indicating higher than scaled solar values of carbon \citep[compare the study of][]{apogee_c}.

In Figure \ref{fig:ba_spectra}, we show the synthetic spectra (crimson line) fits to the observed spectra (black circles) of four stars with \teff\, ranging from 3400 K to 3600 K and \feh\, ranging from -0.5 dex to 0.25 dex. The yellow bands in each panel represent the line masks defined for the Ba line, wherein SME fits observed spectra by varying the barium abundance and finds the best synthetic spectra fit by $\chi^{2}$ minimization. The red band denotes the variation in the synthetic spectrum for a difference of $\pm$0.3\,dex in the [Ba/Fe] indicating the line sensitivity to [Ba/Fe]. The green line shows the synthetic spectrum without Ba, also indicating any possible blends in the line. 

The main spectral features in Figure \ref{fig:ba_spectra}, apart from the Ba line, are due to the first overtone transitions of  $^{12}$CO, both $\Delta\nu = 2\leftarrow 0$ and $\Delta\nu = 3\leftarrow1$ transitions.  The  $^{12}$CO($\Delta\nu = 3-1$) band-head lies just outside the figure at 23220\,\AA. The far blue wing of the CO($\Delta\nu = 3-1$) R63 line to the right of the Ba line marginally influences the Ba line.  This influence is taken care of well since this CO line is modelled well, as are the other high rotational transitions of the CO($\Delta\nu = 3-1$) transition in the Figure.  These lines are formed deeper in the atmosphere than the  lower rotational transitions, i.e. the stronger lines in the Figure (CO($\Delta\nu = 3-1$) R37), which are not properly modelled. Lines with lower excitation energies form further out where the modeling of the atmosphere is less certain and 3D and non-LTE effects may play a large role. This was indeed also seen in the H-band study of high resolution spectra by \citet{montelius:22}, discussing blending CO lines. 


The orange line shows the telluric spectrum of the standard star that is used to correct for telluric contaminations in the observed star spectrum. The telluric spectra are divided out and the telluric lines should have ideally been removed, resulting in a pure stellar spectrum, although with slightly larger noise. However, the telluric division is not always perfect, so the purpose of showing the orange telluric spectra is to show where we might expect spurious spectral features which are actually residuals from the telluric reduction procedure. 

The [Ba/Fe] value corresponding to the best-fit case for the Ba\,{\sc i} line are listed in each panel. We have scaled the barium abundances with respect to solar abundance value of 2.17 from \cite{solar:sme}. It is clear from Figure \ref{fig:ba_spectra} that the Ba \,{\sc i} line at 23253.56 \AA\, is independent of any atomic or molecular line blends, is sensitive to barium abundance and is strong enough for the temperature and metallicity ranges of stars in our sample and thus ensure that the barium abundances determined in this work is reliable.

\begin{table*}[h!]
\caption{Atomic line data for the Ba\,{\sc i} 6s5d$\,^3$D $\rightarrow$ 6s6p\,$^3$P$^o$ fine structure transitions.}
\label{table:transitions}
\centering
\begin{tabular}{c c c c c c c c}
\hline
Lower Term & Upper Term & Wavelength$^a$  & Excitation &log $gf^b$ &  Rel. Int.$^a$ &  Rel. line strength  & comment\\
 & & in Air (\AA) &  Energy (eV) & & & $gf\cdot e^{\chi_\mathrm{exc}/\mathrm{k}T}$ \\
\hline
6s5d $^3$D$_1$ & 6s6p $^3$P$^o_2$ & 22\,311.47 & 1.120  & -2.307 &560   & 0.013 & too weak \\
6s5d $^3$D$_2$ & 6s6p $^3$P$^o_2$ & 23\,253.56 & 1.143  & -1.172 &9900  & 0.19  & observed\\
6s5d $^3$D$_3$ & 6s6p $^3$P$^o_2$ & 25\,514.88 & 1.190  & -0.460 &10000 & 1 &  beyond IGRINS coverage\\
6s5d $^3$D$_1$ & 6s6p $^3$P$^o_1$ & 27\,749.72 & 1.120  & -1.229 &blend & 0.14 & between K and L bands\\
6s5d $^3$D$_2$ & 6s6p $^3$P$^o_1$ & 29\,222.20 & 1.143  & -0.791 &213   & 0.44 & between K and L bands\\
6s5d $^3$D$_1$ & 6s6p $^3$P$^o_0$ & 30\,931.59 & 1.120  & -1.149 &402   & 0.19 & L band\\
\hline 
\end{tabular}
$^{a}$ Measured by \citet{Ba:99} and reported in the NIST ADS database \citep{NIST}.\\
$^{b}$ Oscillator strengths as given by \citet{Kurucz2017}
\end{table*}

\section{Results}
\label{sec:results}

\begin{figure}
  \includegraphics[width=\columnwidth]{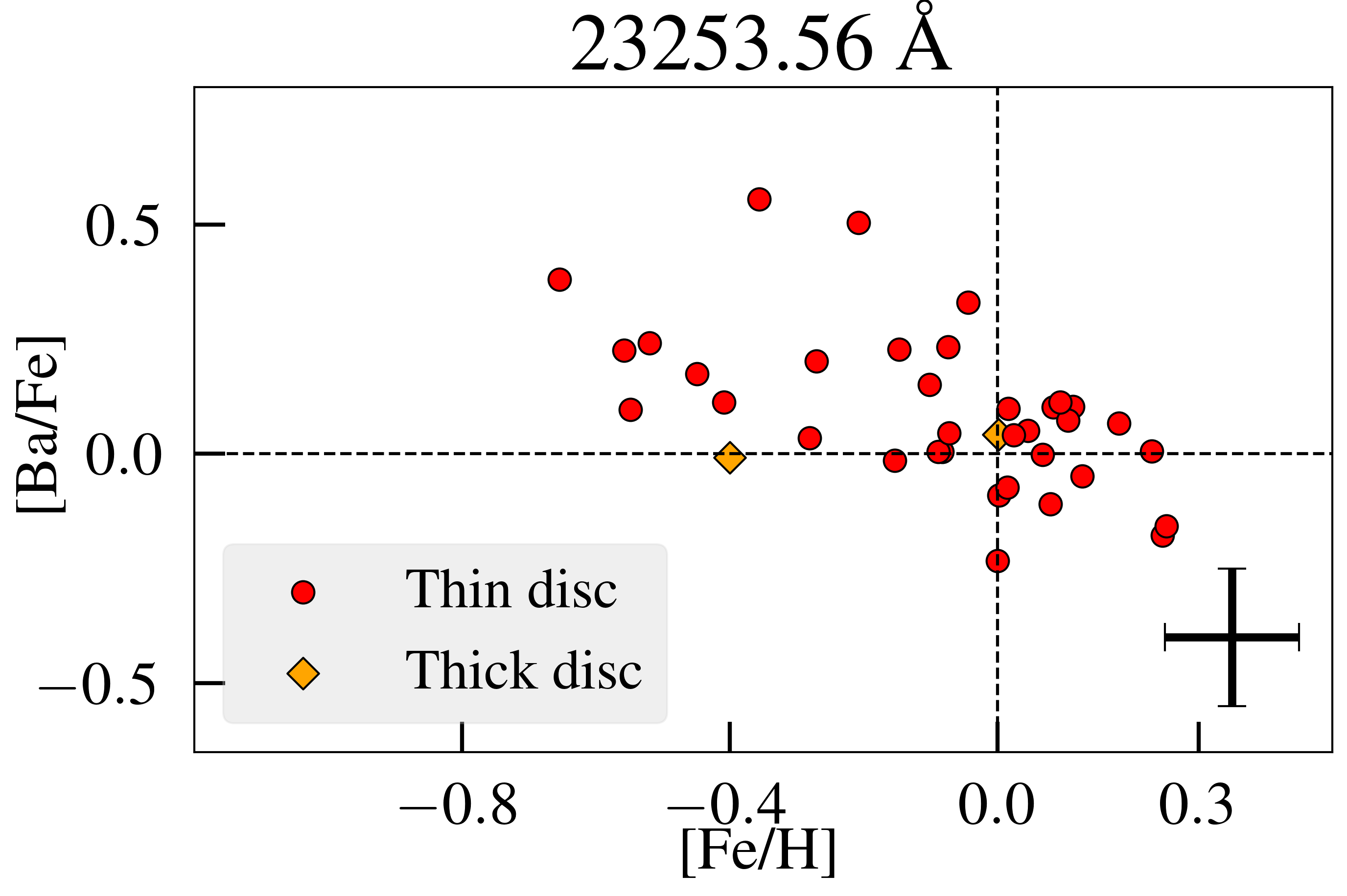}
  \caption{[Ba/Fe] vs. [Fe/H] for stars in our sample with their respective stellar population represented by different symbols (thin disk -
red circle, thick disc - orange diamond). }
  \label{fig:ba_trend}%
\end{figure}


Figure \ref{fig:ba_trend} shows the barium abundance trend as a function of metallicity for the 37 stars in our sample. We see a general downward trend with metallicity, with a larger scatter between metallicities above \feh$>-0.5$ up to solar metallicity, typical for s-process elements. We also find s-element enhanced stars at subsolar metallicities. 

We followed the method described in \citetalias{Nandakumar:2023_I} and \citetalias{Nandakumar:24_III} to determine uncertainties in barium abundances arising from typical uncertainties in the stellar parameters ($\pm$100 K in \teff, $\pm$0.2 dex in \logg, $\pm$0.1 dex in \feh, and  $\pm$0.1 \kms\ in $\xi_\mathrm{micro}$). For this, we chose the metal-rich star 2M14261117–6240220 and redetermined 50 values of the barium abundances by setting the stellar parameters to randomly chosen values from normal distributions, with  the actual stellar parameter value as the mean and the typical uncertainties as the standard deviation. The uncertainty in the barium abundance was then taken to be the mid-value of the difference between the 84$^\mathrm{th}$ and 16$^\mathrm{th}$ percentiles of the abundance distribution, which is estimated to be $\pm$ 0.1 dex. We further estimated uncertainty of $\sim$ 0.1 dex due to uncertainties in fitting the neighboring CO lines and $\sim$ 0.05 dex due to the wrong selection of continuum points surrounding the barium line. Adding all these uncertainties in quadrature, we estimated a typical uncertainty of 0.15 dex which is indicated as the error bar in Figure \ref{fig:ba_trend}. In order to retrieve an accurate barium abundance, the blending CO line wing has to be taken into account and modelled correctly.

Reliable barium abundance has been estimated only for one thick disc star at sub-solar metallicity and it is found to be lower than that for thin disc stars at similar metallicities. This fits well in the overall picture of neutron-capture elements in the disk components found in \citet{DelgadoMena:2017,taut:21,Rebecca_phd}, with the thick-disk following the lower envelope of the thin-disk "cloud". This is also found in the trends of other s-process elements such as cerium and neodymium in \citetalias{Nandakumar:24_III}. Similar observations have been made by \cite{Liu:2020} in their analysis of high-resolution, high S/N spectra of 602 stars from the Keck/HIRES finding the mean value of [Ba/Fe] for the high-$\alpha$ (thick disc) stars to be lower compared to the low-$\alpha$ (thin disc) stars. At the same time, \cite{Prantzos:2023} have shown using chemical evolution models that the thin and thick disc trends overlap and follow a single branch behaviour in the [Ba/Fe] vs [Fe/H] diagram. It is crucial to observe more thick disc stars and carry out more accurate analyses in a larger sample of stars to answer this question.

[Ba/Fe] values estimated for the 37 stars in this work are listed in Table \ref{table:parameters}. We scaled the abundances with respect to solar abundance values from \cite{solar:sme}.

\begin{figure*}
  \includegraphics[width=\textwidth]{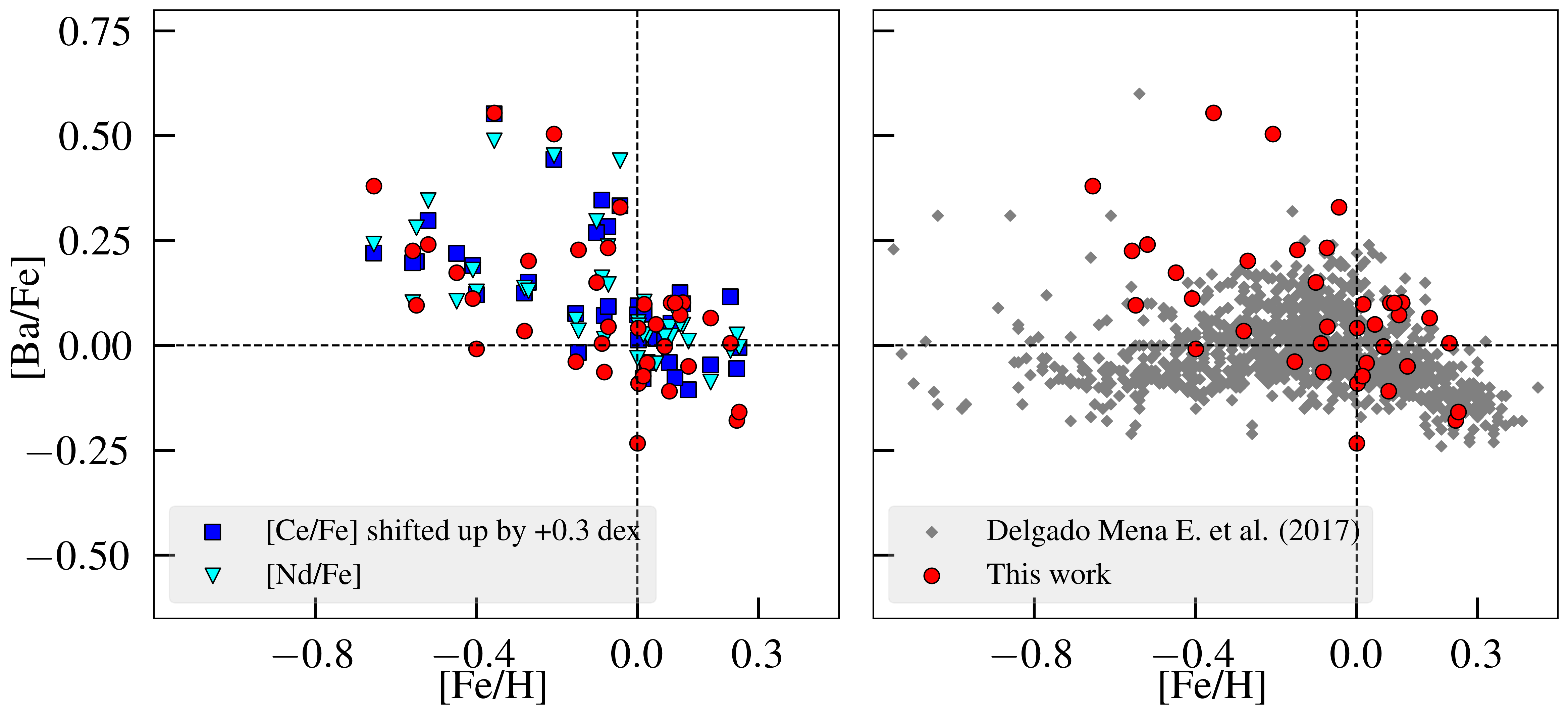}
  \caption{Comparison of barium abundance trends in this work with other s-process elements such as cerium ([Ce/Fe]) and neodymium ([Nd/Fe]) for the same stars (left panel) and with literature values in \cite{DelgadoMena:2017} (right panel).}
  \label{fig:ba_lit}%
\end{figure*}

\section{Discussion}
\label{sec:discussion}

In this section we discuss the atomic barium transitions available in the near-IR wavelengths, compare with the abundance ratio trends with metallicity of barium in the literature and of other s-process elements, derived for the same stars from the same spectra. Finally we discuss in which stars the barium line is detectable and useful for an abundance analysis.

\subsection{The fine-structure lines of barium in the K-band}
\label{sec:atomic}
\begin{figure}
  \includegraphics[width=0.45\textwidth]{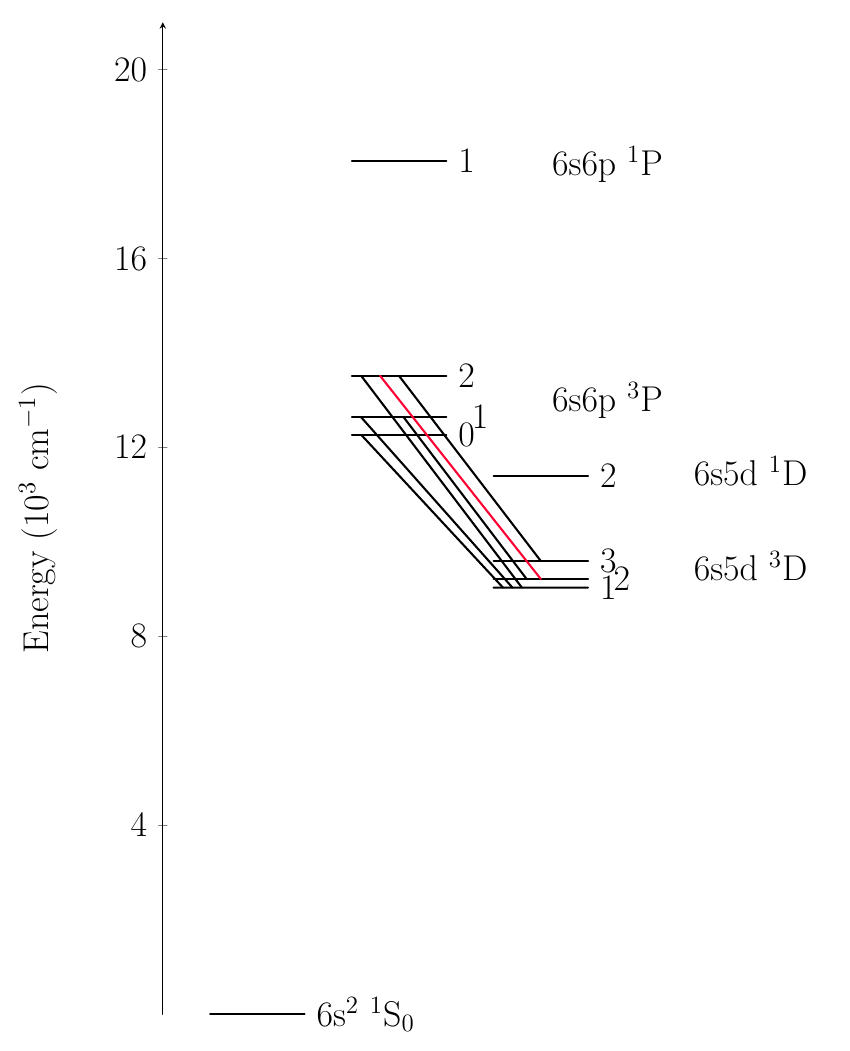}
  \caption{Energy level diagram showing the lowest twelve energy levels in neutral barium, Ba\,{\sc i}. The line used in this study, 6s5d~$^3$D$_2$ -- 6s6p~$^3$P$_2$, is marked in red.}
  \label{fig:ba_grotrian}%
\end{figure}

The barium line identified is a 6s5d\,$^3$D$_2$ -- 6s6p\,$^3$P$^o_2$ transition in Ba\,{\sc i}, with a lower  excitation level at 1.14\,eV, see Table\,\ref{table:transitions}. 
Neutral barium is a two-electron system, and the lower part of the energy level diagram is shown in Figure \ref{fig:ba_grotrian}. For near-infrared studies, transitions from 6s5d to 6s6p levels are expected to be the strongest lines in the spectrum, as the lower $6s5d$ states have excitation energies of only $\sim1.1$\,eV and the transition from the ground state to these levels are forbidden (same parity).

The 6s5d-6s6p multiplet has six lines, as is marked in Figure \ref{fig:ba_grotrian} and the atomic data detailed in Table\,\ref{table:transitions}. Two lines fall in the $K$-band, where the 6s5d $^3$D$_2$ -- 6s6p $^3$P$^o_2$ transition at 23\,253.56\,\AA\ is the strongest one, with 1.1 dex larger log~$gf$ 
The four transitions at longer wavelengths fall outside the $K$-band. 

The wavelengths of the near-IR transitions are measured by \citet{Ba:99} using archive spectra from the Kitt Peak Fourier transform spectrometer. In addition, they provide line intensities, but these are not corrected for instrumental transmission and should thus be treated with care. These wavelengths and intensities are included in the NIST ADS database \citep{NIST} and are given in Table\,\ref{table:transitions}. In the  Table, we also calculate the relative line strengths, calculated with the weak-line formula, where the equivalent width is given by the proportionality $$gf\cdot e^{\chi_\mathrm{exc}/\mathrm{k}T}.$$ From this we see that the line at 22311.47 \AA\ is more than 10 times weaker than the detected line at 23253.56 \AA\, which explains why the 22311.47 \AA\ line is not detected in our spectra of the K band.

There exist no experimental oscillator strengths for the near-IR 6s5d - 6s6p transition multiplet. The most recent theoretical values are from semi-empirical fitting by \citet{Kurucz2017}, and these are added to Table\ref{table:transitions}. 

\begin{figure}
  \includegraphics[width=0.5\textwidth]{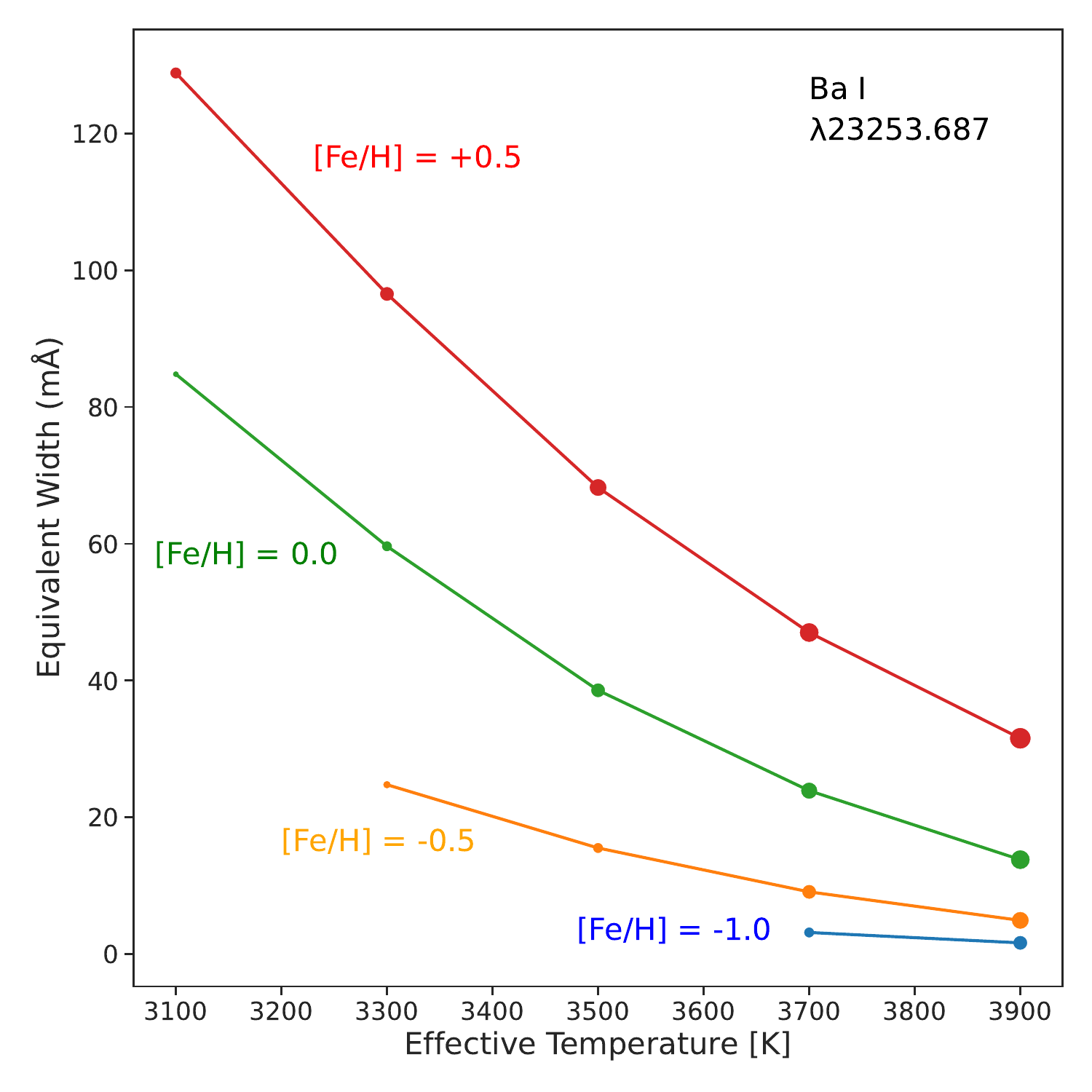}
  \caption{ Equivalent widths in  m\AA\ of the Barium line for typical red giants as a function of temperature. The coloured lines show equal metallicities (blue for \feh\,$=-1.0$, orange for \feh\,$=-0.5$, green for \feh\,$=0.0$, and red for \feh\,$=+0.5$). The size of the symbols indicate the surface gravities corresponding to a given \teff\ and \feh\  according to a typical isochron. The sizes go from largest being \logg\,$= 1.7$ to the smallest being \logg\,$= 0.0$. The equivalent width of the Ba line is quite insensitive to the surface gravity (a change of 1 dex in \logg changes the equivalent width with less then 3\%.).}
  \label{fig:ba_eqw}%
\end{figure}

With an ionization potential of only 5.2\,eV \citep{Ba:99}, most Ba is ionized and Ba\,{\sc i} is a minority species. The line opacity is,  therefore, sensitive to the electron pressure as is the continuous opacity, due to H$^-_\mathrm{ff}$. In general, strengths of weak lines depend on the line and continuum opacities, $\chi_\mathrm{line}$ and $\chi_\mathrm{cont}$, such that the equivalent width of a line $W \propto \chi_\mathrm{line}/\chi_\mathrm{cont}$.  This means that the sensitivity to the electron pressure cancels out in this ratio, and therefore line strength, to a certain degree.  This explains the very small changes in line strengths for varying \logg\ (related to the electron pressure thought the hydrostatic equilibrium equation) shown later in Figure \ref{fig:ba_eqw}.

A previous set of data for Ba\,{\sc i} is estimated from laboratory spectra, as given by \citet{Kurucz93} and included in the VALD database \citep{vald,vald5}. This data is, however, not consistent with the more recent data from \citet{Kurucz2017} or by the experimental data by \citet{Ba:99}. A particular example is the two lines in the K-band region, where 22311.47 and 23253.56 \AA\ lines are predicted to be of similar strengths in the earlier data, whereas the latter data is consistent with our findings that the 23253.56 \AA\ line is observed but the 22311.47 \AA\ line is much weaker. 






\subsection{Comparison with other s-process elements and literature}
\label{sec:ba_trend_comparison}

In \citetalias{Nandakumar:24_III}, abundances of other s-process elements have been determined for the same stars analysed in this work. Hence it is only logical to compare the barium abundance trends determined in this work with other s-process element abundances for the same stars. In the left panel of Figure \ref{fig:ba_lit}, we compare the abundance trends of cerium (blue squares) and neodymium (cyan inverted triangles) with the barium abundance trends (red circles) in this work. The cerium abundances have been shifted by +0.3 dex to pass through the solar value at solar metallicity. Both the cerium and neodymium abundances exhibit very similar trends as already seen in the mean s-process abundance plot (Figure 23) in \citetalias{Nandakumar:24_III}. The barium abundances in this work has larger scatter but show a qualitatively similar trend as those for the cerium and neodymium abundances. We note that the four stars that have high mean s-process abundances in \citetalias{Nandakumar:24_III} also show higher barium abundances. This is especially evident in the case of the two sub-solar metallicity stars, 2M06520463-0047080 and 2M06574070-1231239, with similar high abundances in cerium and neodymium. This is further evidence that we have been able to determine reliable abundances from the barium line at 23253.56 \AA.

In order to compare our barium abundance trends with metallicity with those from the literature we chose the barium abundances determined from high resolution (R $\sim$ 115 000) optical spectra (4000 - 7000 \AA) of 1111 FGK dwarf stars from the HARPS GTO planet search program in \cite{DelgadoMena:2017}. We show this comparison in the right panel of the Figure \ref{fig:ba_lit}.  Our barium abundances generally lie within the overall trend of \cite{DelgadoMena:2017}, especially at metallicities above -0.5 dex. Our trend do not exhibit the decrease in barium abundances from super-solar to sub-solar values at metallicities below -0.5 dex as evident in the case of the \cite{DelgadoMena:2017} trend. We have only four stars with reliable barium measurements at metallicities below -0.5 dex, and more stars at lower metallicities are needed to confirm this trend using the abundance measured from the barium line at 23253.56 \AA. At the same time, we note that there are stars with similar high barium abundance measurements at similar lower metallicities in \cite{DelgadoMena:2017} as well.

\subsection{Best stars to measure Ba\,{\sc i} line at 23253.56 \AA}
\label{sec:eqw}

In Figure \ref{fig:ba_spectra}, we have shown that the Ba\,{\sc i} line at 23253.56 \AA\, is devoid of any atomic or molecular blends making it reliable to measure abundances from. Furthermore, in the previous sections, we have demonstrated that the barium abundances trend determined from the Ba\,{\sc i} line at 23253.56 \AA\, is qualitatively consistent with other s-process elemental abundance trends determined for the same stars and also with reliable barium abundance trends in the literature. We also find high barium abundances for the two stars with high cerium and neodymium abundances. 

Considering that there is a lack of useful spectral lines in the near-IR for determining the abundances of neutron-capture elements, we want to investigate which type of stars should be targeted in the future to get reliable barium abundances from this Ba\,{\sc i} line. In order to investigate this, we estimate the equivalent widths of the Ba\,{\sc i} line for typical red giant stars with \teff\, from 3100 K to 3900 K (in 200 K intervals), \logg\, from 0.0 to 1.7 dex, and \feh\, values of -1.0, -0.5, 0.0, and +0.5 dex. In Figure \ref{fig:ba_eqw}, we show the resulting equivalent width measurements for different combinations of \teff, \logg\ and \feh.  The coloured lines show different metallicities: blue for \feh\,= -1.0, orange for \feh\,= -0.5, green for \feh\,= 0.0, and red for \feh\,= +0.5. The size of the symbols indicate the surface gravities corresponding to a given \teff\, and \feh\, according to a typical isochrone. The sizes go from the largest being \logg\,= 1.7 to the smallest being \logg\,= 0.0. 

From Figure \ref{fig:ba_eqw}, it is evident that the Ba\,{\sc i} line at 23253.56 \AA\, is the strongest for cool (\teff\,$<$ 3500 K), solar and super-solar metallicity red giant stars. Even though the line gets weaker with increasing \teff\, and decreasing \feh, we have demonstrated that it is possible to determine reliable abundances from this line in high resolution, high S/N spectra acquired with a capable instrument such as IGRINS. But at very low metallicities (\feh $\sim$ -1.0), this barium line will be very weak making it impossible to determine abundances. We also note that the equivalent width of the Ba line is quite insensitive to the surface gravity (a change of 1 dex in \logg\, changes the equivalent width with less then 3\%.), which is indeed expected as described in Section \ref{sec:atomic}.

\section{Conclusions}
\label{sec:conclusion}

In this work, we have identified and characterised a Ba\,{\sc i} line at 23253.56 \AA\, in  high-resolution (R$\sim$45 000) IGRINS spectra of M giants (\teff$<$4000 K) in the solar neighborhood. While there are five other Ba\,{\sc i} lines within the same fine structure multiplet, the line at 23253.56 \AA\, is found to be the strongest and  with the highest transition probability in the K-band (see Section \ref{sec:atomic}). 

We have shown in Figure \ref{fig:ba_spectra} that the Ba\,{\sc i} line at 23253.56 \AA\, is devoid of any atomic or molecular line blends, is sensitive to barium abundances and strong enough in the spectra of the stars in our sample to determine reliable barium abundances from. We determined [Ba/Fe] for 37 M giants and found a general downward trend with metallicity, with a larger scatter between metallicities above \feh$>$ -0.5 up to solar metallicity, typical for s-process elements (Figure \ref{fig:ba_trend}). 

We have shown that our [Ba/Fe] versus [Fe/H] trend is very similar to the other s-process element trends (cerium and neodymium) determined for the same stars from the same spectra. The two barium enhanced stars with similar high cerium and neodymium abundances further confirms this similarity. We further demonstrated that our [Ba/Fe] versus [Fe/H] trend, within uncertainties, is consistent with barium abundance trends in the literature. Our investigations with the equivalent width measurements of synthetic spectra of stars with different combinations of \teff, \logg, and \feh\, indicate that the Ba\,{\sc i} line at 23253.56 \AA\, is the strongest for cool (\teff\,$<$ 3500 K), solar and super-solar metallicity giant stars and becomes gradually weaker with decreasing \teff\, and \feh. 

Thus, in this work, we have successfully identified and characterized the Ba\,{\sc i} line at 23253.56 \AA\ and determined reliable barium abundances for 37 M giants from their high resolution, high S/N IGRINS spectra. We also emphasize the importance of systematic and consistent analysis in the determination of stellar parameters and abundances to avoid systematic effects. This brings forth the importance of a capable near-IR instruments such as IGRINS in the identification and characterization of unidentified atomic and molecular lines of rare and interesting chemical species in the near-IR wavelength regime \citep[see also][]{Cunha:2017,montelius:22,nandakumar:22_P,Nandakumar:2023b_II}. This work is thus one of the prime examples of the immense possibilities of the existing (GIANO, CRIRES+), near-future (MOONS), and future high-resolution near-IR spectrometers (proposed for the extremely large telescopes such as the ELT and TMT) as they will enable observing more distant M giants, including in other galaxies.

\begin{acknowledgements}
We thank the anonymous referee for the constructive comments and suggestions that improved the quality of the paper. G.N.\ acknowledges the support from the Wenner-Gren Foundations (UPD2020-0191 and UPD2022-0059) and the Royal Physiographic Society in Lund through the Stiftelsen Walter Gyllenbergs fond. G.N.\ also acknowledges the support received from the Royal Swedish Academy of Sciences. Vetenskapsakademiens stiftelser. N.R.\ acknowledge support from the Swedish Research Council (grant 2023-04744) and the Royal Physiographic Society in Lund through the Stiftelsen Walter Gyllenbergs and Märta och Erik Holmbergs donations. H.H.\ acknowledges support from the Swedish Research Concil VR (grant 2023-05367). 
This work used The Immersion Grating Infrared Spectrometer (IGRINS) was developed under a collaboration between the University of Texas at Austin and the Korea Astronomy and Space Science Institute (KASI) with the financial support of the US National Science Foundation under grants AST-1229522, AST-1702267 and AST-1908892, McDonald Observatory of the University of Texas at Austin, the Korean GMT Project of KASI, the Mt. Cuba Astronomical Foundation and Gemini Observatory.
This work is based on observations obtained at the international Gemini Observatory, a program of NSF’s NOIRLab, which is managed by the Association of Universities for Research in Astronomy (AURA) under a cooperative agreement with the National Science Foundation on behalf of the Gemini Observatory partnership: the National Science Foundation (United States), National Research Council (Canada), Agencia Nacional de Investigaci\'{o}n y Desarrollo (Chile), Ministerio de Ciencia, Tecnolog\'{i}a e Innovaci\'{o}n (Argentina), Minist\'{e}rio da Ci\^{e}ncia, Tecnologia, Inova\c{c}\~{o}es e Comunica\c{c}\~{o}es (Brazil), and Korea Astronomy and Space Science Institute (Republic of Korea).
The following software and programming languages made this
research possible: TOPCAT (version 4.6; \citealt{topcat}); Python (version 3.8) and its packages ASTROPY (version 5.0; \citealt{astropy}), SCIPY \citep{scipy}, MATPLOTLIB \citep{matplotlib} and NUMPY \citep{numpy}.
\end{acknowledgements}

%
%


\bibliographystyle{aa}
\bibliography{references} 







\end{document}